\documentclass[a4paper,11pt]{article}
\usepackage{jheppub}

\usepackage{booktabs}
\usepackage{braket}
\usepackage{graphicx}
\usepackage{slashed}
\usepackage[dvipsnames]{xcolor}
\def \Br{\textrm{Br}}

\def \GeV{\textrm{GeV}}
\def \ps{\textrm{ps}}

\def \MSbar{\overline{\textrm{MS}}}
\allowdisplaybreaks

\newcommand{\lmu}{l_\mu}
\newcommand{\lc}{l_{m_c}}
\title{\boldmath 
The Standard Model Prediction for the Rare Decay  $B \to X_s \nu \bar \nu$}
\author[a,b]{Matteo Fael}
\author[c]{Jack Jenkins}
\author[d]{Enrico Lunghi}
\author[e,f]{Zachary Polonsky}

\affiliation[a]{
Dipartimento di Fisica e Astronomia ``G.~Galilei,'' Università di Padova, Via F.~Marzolo 8, 35131
Padova, Italy
}
\affiliation[b]{
Istituto Nazionale di Fisica Nucleare, Sezione di Padova, Via F.~Marzolo 8, 35131 Padova, Italy
}
\affiliation[c]{
Theoretische Physik 1, Center for Particle Physics Siegen (CPPS), Universit\"{a}t Siegen,
Walter-Flex-Stra\ss e 3, D-57068 Siegen, Germany
}
\affiliation[d]{
Physics Department, Indiana University, Bloomington, IN 47405, USA
}
\affiliation[e]{
Physik-Institut, Universit\"at Z\"urich, CH-8057 Z\"urich, Switzerland
}
\affiliation[f]{
Department of Physics, Jo\v{z}ef Stefan Institute, 1000 Ljubljana, Slovenia
}

\emailAdd{matteo.fael@pd.infn.it}
\emailAdd{Jack.Jenkins@uni-siegen.de}
\emailAdd{elunghi@iu.edu}
\emailAdd{zpolonsky@gmail.com}

\date{April 2025}
\abstract{
We present updated and comprehensive Standard Model predictions for the inclusive rare decay $B \to X_s \nu\bar{\nu}$. Using a state-of-the-art determination of the short-distance coefficient, including NLO QCD and electroweak effects, and implementing a consistent treatment of heavy-quark masses and power corrections within the kinetic scheme, we compute the total decay rate and the neutrino invariant mass spectrum. We incorporate all known perturbative contributions in the heavy quark limit up to $\mathcal{O}(\alpha_s^3)$, evaluate non-perturbative corrections through the Heavy Quark Expansion up to $1/m_b^3$, and include estimates of four-quark operator matrix elements from HQET sum rules. We obtain the Standard Model branching ratios
$
\text{Br}(B^0 \to X_s\nu\bar{\nu}) = (3.35 \pm 0.10)\times 10^{-5}
$
and
$
\text{Br}(B^+ \to X_s\nu\bar{\nu}) = (3.62 \pm 0.11)\times 10^{-5},
$
representing a significant improvement in both central value and precision compared to previous determinations. We also provide predictions for partial rates with kinematic cuts relevant to Belle II. Our results are timely in view of recent measurements of $B \to K\nu\bar{\nu}$ and the first upper limits on inclusive $B \to X_s\nu\bar{\nu}$, and they offer a robust baseline for future searches for physics beyond the Standard Model in $b\to s\nu\bar{\nu}$ transitions.
}

\preprint{
\begin{minipage}{3cm}
\small
\flushright
P3H-25-113 \\
SI-HEP-2025-32
\end{minipage}}

\begin{document}

\maketitle

\section{Introduction}
Flavor changing neutral currents are prohibited at tree level in the Standard Model (SM) by the unitarity of the Cabibbo-Kobayashi-Maskawa (CKM) matrix~\cite{Cabibbo:1963yz, Kobayashi:1973fv} and, therefore, offer an excellent opportunity to observe new physics effects. At the loop level, the GIM mechanism~\cite{Glashow:1970gm} strongly favors $b\to s$ transitions whose amplitudes scale with the relatively large combination $V_{ts}^{\,*}V_{t\smash{b}}^{\vphantom{*}}\times f(m_t^2/m_W^2) \sim O(10^{-2})$. Among all rare processes involving a $b$-hadron, decays induced by the quark level transitions $b\to s (\gamma, \ell^+ \ell^-, \nu\bar\nu)$ are the most promising because they do not suffer from the very large theoretical uncertainties that plague fully hadronic modes.

Recently the Belle II collaboration presented a measurement of the exclusive decay $B^+ \to K^+ \nu\bar\nu$~\cite{Belle-II:2023esi, Belle-II:2025lfq},  which deviates from the SM prediction at the 2.7$\sigma$ level, and an upper limit on the fully inclusive $B\to X_s\nu\bar\nu$ rate, using full event reconstruction for a rare semileptonic decay for the first time, which is about an order of magnitude above the SM prediction (prior to this measurement, BaBar~\cite{BaBar:2013npw} and Belle~\cite{Belle:2005nvz, Belle:2006mdr, Belle:2017oht} presented upper limits on the $B\to K^{(*)} \nu\bar\nu$ branching ratios). Prompted by these results, in this paper we present the state-of-the-art SM prediction for this inclusive decay rate. Our main result for the isospin averaged rate and the Belle II upper limit~\cite{Belle-II:2025bho} read:
\begin{align}
\text{Br} (B\to X_s \nu\bar\nu)_{\rm exp} \; & < \; 3.2 \times  10^{-4} \; \text{at 90\% C.L.}\; , \label{eq:BRexp}\\
\text{Br} (B\to X_s \nu\bar\nu)_{\rm SM} \; & = \; (3.48 \pm 0.11) \times 10^{-5} \; . \label{eq:BRSMiso} 
\end{align}
Currently the Belle II upper limit is limited by systematic uncertainties as can be seen from the result of their branching ratio $\text{Br} (B\to X_s \nu\bar\nu)_{\rm exp} \allowbreak = [ 8.8^{+8.5}_{-8.2}{\rm (stat)}^{+12.6}_{-10.8}{\rm (syst)}] \allowbreak \times 10^{-5}$. The two dominant sources of systematics are the simulated-sample size and background normalization, which will be hopefully reduced, together with statistical uncertainties, in future updates. 

From a theoretical point of view, these decays are extremely interesting for their direct implications to dark matter and axion searches (see, for instance, Refs.~\cite{Berezhnoy:2025osn, Gao:2025ohi} for recent studies), and for their potential connection to the famous anomalies in global $b\to s \ell^+\ell^-$ fits (see, for instance, Refs.~\cite{Capdevila:2023yhq, Hurth:2025vfx} for recent updates). The latter, as it is well known, are plagued by potentially large uncertainties associated to charm loops. In exclusive $B\to K^{(*)} \ell^+ \ell^-$ decays, charm effects appear in certain non-local power corrections (the so-called charming penguins), which are currently impossible to calculate and whose size is currently the subject of a very vibrant debate within the community (see, for instance, Refs.~\cite{Gubernari:2020eft, Ciuchini:2021smi, Gubernari:2022hxn, Isidori:2024lng, Isidori:2025dkp, Frezzotti:2025hif}). Inclusive $B\to X_s \ell^+\ell^-$ are controlled by the same underlying operators but charm loops are under much better control: factorizable charmonium rescattering can be extracted from $e^+e^-\to {\rm hadrons}$ data via a dispersion relation while non-factorizable effects are described in terms of resolved photon contributions and local power corrections at low and high dilepton invariant mass, respectively (see, for instance, Refs.~\cite{Huber:2020vup, Huber:2024rbw}). In contrast, both exclusive and inclusive $b\to s \nu\bar\nu$ decays are completely insensitive to charm rescattering effects. The only issue which the $\nu\bar\nu$ and $\ell^+\ell^-$ inclusive modes share is the need for an extrapolation in $M_{X_s}$ from the fiducial region $M_{X_s} \lesssim 2 \; \text{GeV}$ to the whole phase space~\cite{Huber:2023qse} which is currently performed with the aid of a one-parameter Fermi motion model~\cite{Belle-II:2023esi}.

Another interesting question is whether it is possible to connect New Physics contributions to $b\to s\ell^+ \ell^-$ and $b\to s\nu\bar\nu$. While explicit new physics models tend to yield contributions to both modes (see, for instance, Ref.~\cite{Altmannshofer:2009ma, Buras:2014fpa}), it is interesting to study the question from the SMEFT point of view. Following the analysis presented in Ref.~\cite{Grunwald:2023nli} (see also Ref.~\cite{Cornella:2021sby, Isidori:2023pyp, Chen:2024jlj, Allwicher:2023xba, Das:2023kch, Allwicher:2024ncl}) one can easily write the contributions to the two relevant operators to $b\to s\ell\ell$ ($O_9$ and $O_{10}$) and to the single operator responsible for $b\to s\nu\bar\nu$ ($O_L$) in terms of SMEFT coefficients (assuming lepton flavor universality and minimal flavor violation):
\begin{align}
\Delta C_9 & \propto \; \tilde C_{lq}^+ + (4 s_W^2 -1) \tilde C_{\varphi q}^+ + \tilde C_{qe}  \simeq \; \tilde C_{lq}^+ + \tilde C_{qe} , \\
\Delta C_{10} & \propto \; - \tilde C_{lq}^+ + \tilde C_{\varphi q}^+ + \tilde C_{qe} \; , \\
\Delta C_{L} & \propto \;  \tilde C_{l q}^- + \tilde C_{\varphi q}^+ \; .
\end{align}
Assuming that the $b\to s\ell^+\ell^-$ anomalies are due to a large negative shift in $C_9$ {\it without} a corresponding contribution to $C_{10}$ (as is currently suggested by $\bar{B}_s \to \mu^+ \mu^-$ in the global fits), we see that either $\tilde C_{lq}^+$ or $\tilde C_{qe}$ must be large and negative. In the former case, a corresponding negative contribution to $\tilde C_{\varphi q}^+$ is required to suppress $\Delta C_{10}$, thus we expect a large and {\it negative} contribution to $\Delta C_L$; in the latter case, the large and negative contribution to $\tilde C_{qe}^+$ requires a positive $\tilde C_{\varphi q}^+$, thus inducing a large and {\it positive} contribution to $\Delta C_L$. Only a situation in which the contribution to $\Delta C_9$ is shared between $\tilde C_{lq}^+$ and $\tilde C_{qe}^+$ allows the cancellation of the effect on $\Delta C_{10}$ without an accompanying large $\tilde C^+_{\varphi q}$. We conclude that the correlation between the $\ell\ell$ and $\nu\bar\nu$ modes, while not guaranteed, is certainly a possibility. 

The paper is organized as follows. In section~\ref{sec:HEW}, we introduce the effective Hamiltonian and the scale-invariant Wilson coefficient $\tilde X_t$, which we evaluate including next-to-leading-order (NLO) QCD and EW corrections. In particular we discuss the uncertainty on the top quark mass and the impact that the inclusion of higher order terms in its running has on the residual scale dependence of $\tilde X_t$. In section~\ref{sec:ME} we discuss the matrix elements of the dimension 5 and 6 operators which appear in $1/m_b^{2}$ and $1/m_b^{3}$ power corrections. In section~\ref{sec:TotalRate}, we present the total $ B \to X_s \nu\bar\nu$ rate which we obtain combining the NLO Wilson coefficient $\tilde X_t$ with the next-to-next-to-next-to-leading-order (N3LO) calculation of the required matrix elements (which is identical to the $B \to X_u \ell\nu$ case). The combination of NLO and N3LO corrections to Wilson coefficient and matrix elements is permissible because of the scale invariance of $\tilde X_t$. In section~\ref{sec:kinetic}, we express the total rate calculated in the pole scheme in terms of short distance masses and HQET matrix elements. In particular, we adopt the kinetic scheme for all bottom related quantities and the MSbar scheme for the charm mass. In section~\ref{sec:spectrum} we present the next-to-next-to-leading-order (NNLO) results for the differential decay rate integrated with a cut on $m^2_{\nu\nu}$. In section~\ref{sec:results} we present the results of a phenomenological analysis based on the improvements discussed above and a comparison with previous studies of these decays. Finally, in section~\ref{sec:conclusions} we summarize our main results and discuss future theoretical and experimental prospects.

\section{Electroweak Hamiltonian}
\label{sec:HEW}
The $B \to X_s \nu \bar{\nu}$ amplitude is described in the SM by an effective $b \to s \nu \bar{\nu}$ Hamiltonian consisting of a single operator at dimension six 
\begin{align}
    \mathcal{H}_\mathrm{eff} &=\frac{4G_F}{\sqrt{2}}V_{ts}^* V_{tb}C_\nu (\bar{s}_L \gamma_\mu b_L)\textstyle {\sum}_\ell (\bar{\nu}_{\ell L}\gamma^\mu \nu_{\ell L}) \, . \label{eq:EFTdef}
\end{align}
The hadronic current $\bar{s}_L \gamma_\mu b_L$ in Eq.~\eqref{eq:EFTdef} is conserved in both QCD and QED. Therefore the coefficient $C_\nu$ is scale independent above $\mu \sim m_b$, up to the electroweak scale, in mass-independent schemes such as $\MSbar$. Since the FCNC Wilson coefficient is naturally suppressed by a loop factor and $\alpha$, it is conventionally written as
\begin{align}
    C_\nu&= \frac{\alpha(\mu_w)X_t(\mu_w)}{2\pi \sin^2 \theta_W(\mu_w)} = \frac{\alpha(M_Z) \tilde{X}_t}{2\pi \sin^2 \theta_W(M_Z)} \label{eq:Cnu} \, ,
\end{align}
where $X_t$ is the result of the matching of the SM to the LEFT according to the renormalized couplings $\alpha$ and $\sin^2 \theta_W$ defined in the $\MSbar$ scheme in the five- and six-flavor theories respectively and $\mu_w$ is the electroweak matching scale. Considering the electroweak corrections to $X_t$, it is useful to define a scale-independent function $\tilde{X}_t$ by reference to a specific scale as in Eq.~\eqref{eq:Cnu} above.

While the expressions for the LO and NLO QCD corrections to $X_t$ are relatively compact, the expression for the electroweak correction is lengthy, and the large top limit does not reproduce the complete result for arbitrary $M_{W,Z}/m_t$ to an acceptable accuracy. We resort to implementing the complete top mass dependence of the electroweak correction from Eq.~(4.3) of Ref.~\cite{Brod:2010hi} at a fixed scale $\mu_w=M_Z$ (see also the detailed discussion in Ref.~\cite{Brod:2021hsj}).\footnote{We thank Joachim Brod for clarification on the implementation of the electroweak correction.} In practice we implement the perturbative corrections as
\begin{align}
    \tilde{X_t}=X_t^{(0)}[\overline{m}_t(\mu_w)]+\frac{\alpha_s(\mu_w)}{4\pi}X_t^{(1)}[\overline{m}_t(\mu_w), \ln \frac{\mu}{\overline{m}_t(\mu_w)}] + \frac{\alpha(M_Z)}{4\pi} X_t^{(EW)}[\overline{m}_t(\overline{m}_t)] \label{eq:Xtexpand}
\end{align}
with the QCD coupling likewise defined in the five flavor theory.\footnote{The distinction is of higher order than considered in Eq.~\eqref{eq:Xtexpand} since the coefficient functions of $X_t$ do not resum any large QCD logarithms. The explicit logarithm in Eq.~\eqref{eq:Xtexpand} compensates the running of the top mass in the leading term at $O(\alpha_s)$.} 

Note that for consistency of the electroweak scheme, one should not treat $G_F$, $M_H$, $M_Z$, $M_W$, $\alpha$, and $s_W$ all as independent parameters, and the choice made in Ref.~\cite{Brod:2010hi} is to find $M_W$ in terms of $M_Z$ and $M_H$ using the results of Ref.~\cite{Awramik:2003rn}. However, since in our implementation of the electroweak matching, we use the fit function $r_X$ in Ref.~\cite{Brod:2010hi}, which is presented in a fixed scheme, we simply take the reference value of $M_W$ from Table~\ref{tab:inputs} and presume the electroweak scheme dependence is suppressed to a large extent as is stated by the authors of Ref.~\cite{Brod:2010hi}.

In the evaluation of $\tilde X_t$, the top mass is evaluated in the $\overline{\text{MS}}$ scheme with respect to QCD, requiring as an input the value of $\overline{m}_t(\overline{m}_t)$. Unfortunately, the precision of the $\overline{\text{MS}}$ value of the top mass from cross section measurements is not yet competitive with that of the Monte Carlo (MC) mass extracted from $t\bar t$ kinematics, $M_t^{\text{MC}} = 172.56\pm 0.31$ GeV~\cite{ParticleDataGroup:2024cfk}. As such, we opt to use the latter mass to determine $\overline{m}_t(\overline{m}_t)$ using the following strategy: first, we interpret the MC mass as a \textit{low-scale} mass in the four-flavor MSR scheme at $R_{\text{low}}\sim 2$ GeV. Using the Python implementation of the \texttt{REvolver} code~\cite{Hoang:2021fhn}, we then run this low-scale mass up to the top scale, recoupling the bottom and top quarks to the evolution, converting to the $\overline{\text{MS}}$ scheme in the six-flavor theory to find $\overline{m}_t(\overline{m}_t)$.

This process is repeated using the initial low-scale range $R_{\text{low}}\in[1, 3]$ GeV in order to evaluate an uncertainty associated to the residual scale variation in the MSR scheme. Finally, we add in quadrature an additional 500 MeV ``conceptual uncertainty'' associated to the interpretation of the MC mass as a low-scale mass in the MSR scheme, which effectively specifies an error coming from the fact that the showers generated in Pythia are not NLL precise. This gives the final result
\begin{equation}\label{eq:top_mass}
    \overline{m}_t(\overline{m}_t) = 162.87(17)_{\text{MSR}}(30)_{M_t}(50)_{\text{MC}}\text{ GeV} = 162.87(60)\text{ GeV}\,,
\end{equation}
In Eq.~\eqref{eq:top_mass}, ``\text{MSR}'' denotes the uncertainty associated to the variation of the low scale, which arises due to the fact that the relation between the shower-cut-dependent Pythia MC mass and the MSR mass has not yet been determined. The size of this uncertainty reflects the fact that the difference between shower-cutoff schemes are not anticipated to have a large impact on the relation between the MC mass and the MSR mass. Additionally, ``$M_t$'' denotes the parametric uncertainty from the quoted PDG value of $M_t^{\text{MC}}$, while ``$\text{MC}$'' is the conceptual uncertainty arising from identifying $M_t^{\text{MC}}$ as a low-scale mass. The latter is the dominant source of error in $\overline{m}_t(\overline{m}_t)$ due to the fact that one cannot reliably extract the uncertainty from neglecting NLL effects in the parton showers from within Pythia. Furthermore, there is an additional $\pm 90$ MeV uncertainty arising from the parametric uncertainty from $\alpha_s(M_Z)$. This uncertainty is negligible for our purposes in $\overline{m}_t(\overline{m}_t)$, and we include the parametric uncertainty arising from $\alpha_s(M_Z)$ elsewhere in the computation, so in order to avoid double-counting this uncertainty, we neglect this in Eq.~\eqref{eq:top_mass}. Finally, we note that once NLL-precise parton shower are included in top production and decay within Pythia, the conversion from the MC mass to the pole mass can be analytically computed, and the ``MSR'' and ``MC'' sources of error can be combined into a total uncertainty from neglecting NNLO effects -- likely to be very small. 

Functionally, we then treat Eq.~\eqref{eq:top_mass} as an input parameter in the computation of $\tilde{X}_t$. Finally, including the NLO QCD and NLO EW matching contributions~\cite{Misiak:1999yg,Buchalla:1998ba,Brod:2010hi} gives
\begin{equation}\begin{split}\label{eq:Xt}
    \tilde X_t = 1.471 \pm 0.010_{\text{scale}} \pm 0.006_{\text{param}} = 1.471\pm 0.012\,.
\end{split}
\end{equation}
The central value of Eq.~\eqref{eq:Xt} is determined by evaluating Eq.~\eqref{eq:Xtexpand} at the scale $\mu_w~=~M_Z$, and the scale uncertainty is found by varying the matching scale in the range $\mu\in[60,320]$ GeV and taking half the difference between the maximum and minimum values. The running of the QCD coupling and top mass is performed at five-loop order using \texttt{RunDec}~\cite{Herren:2017osy,Chetyrkin:2000yt}.\footnote{To run the top mass in six-flavor QCD, we first run the five-flavor coupling to $\mu_t=\overline{m}_t$ and match to the six-flavor theory there. We verified that other choices of $\mu_t$ had a negligible effect on the final result.}
Parametric uncertainties come from $\overline{m}_t(\overline{m}_t)$, $\alpha_s(M_Z)$, $M_H$, and $M_Z$ in Table~\ref{tab:inputs} below.

The central value of Eq.~\eqref{eq:Xt} agrees with the value $X_t=1.462 \pm 0.017_{\textrm{scale}}$ in Ref.~\cite{Brod:2021hsj} up to a sub-percent difference, while our uncertainty from scale variation is significantly smaller. The sources of these differences can be mostly attributed to our inclusion of higher-order QCD effects in the running of $\alpha_s$ and $\overline{m}_t(\mu_w)$ as well as the fact that we evaluate Eq.~\eqref{eq:Xt} at the scale $\mu_w = M_Z$ in order to remain consistent with $C_\nu$ as defined in Eq.~\eqref{eq:Cnu} instead of averaging the minimum and maximum values of $\tilde X_t$ over the range of $\mu_w\in[60,320]$ GeV. Running $\alpha_s$ and $\overline{m}_t$ at lower orders, and computing the central value in a way consistent with Refs.~\cite{Brod:2010hi,Brod:2021hsj}, we find
$\tilde X_t = 1.461 \pm 0.015_{\text{scale}}\pm 0.006_{\text{param}}$ where the remaining tiny difference is likely due to a different choice of input parameters.

Finally, in order to calculate the scale uncertainty in the branching ratios, we also consider the running of all $\overline{\text{MS}}$ parameters in the definition of $C_\nu$ at the EW scale, including the leading-logarithmic running of $\alpha(\mu_w)$ and $\sin^2\theta_W(\mu_w)$ appearing in Eq.~\eqref{eq:Cnu}. The leading fixed-order logarithms are in principle canceled by similar terms appearing in the full analytic result of $X_t^{(EW)}(\mu_w)$. Since the fit function given in Ref.~\cite{Brod:2010hi} that we implement does not include any explicit scale dependence, this will lead to an artificial residual $\mu_w$-dependence in $C_\nu$ which should be canceled by the two-loop EW matching contributions. To compensate this, we add a term to $X_t^{(EW)}$ which explicitly cancels this leading fixed-order scale dependence
\begin{equation}\label{eq:ewScale}
 \Delta X^{(EW)}_t(\mu_w) = -\Bigg[\beta_e^{(6)} - \frac{1}{\sin^2\theta_W(M_Z)}\Bigg(\frac{\big(1 - \sin^2\theta_W(M_Z)\big)\beta_2}{\sin^2\theta_W(M_Z)} - \beta_1\Bigg)\Bigg]X_t^{(0)}\log\Big(\frac{\mu_w^2}{M_Z^2}\Big) 
\end{equation}
where $\beta_e^{(6)} = \beta_1+\beta_2 = 10$, $\beta_1 = 41/6$, and $\beta_2 = 19/6$ are the one-loop beta functions for $\alpha$ in the six-flavor theory, the weak hypercharge $\alpha_1$, and the weak isospin $\alpha_2$. With this procedure, we obtain $C_\nu= (7.916\pm0.053_{\text{scale}}\pm0.032_{\text{param}})\times 10^{-3}$, which differs in uncertainty from $C_\nu= (7.916\pm0.049_{\text{scale}}\pm0.032_{\text{param}})\times 10^{-3}$ obtained without running the electroweak coupling constants.

\section{Matrix Elements}
\label{sec:ME}
The spectrum and decay rate of $B \to X_s \nu \bar{\nu}$ are identical (up to CKM and electroweak normalization) to that of $B \to X_u \ell \nu$ in the limit $m_s=0$ (which we assume in this work).
Inclusive decays can be described by a local operator product expansion,
the Heavy Quark Expansion (HQE), as the decay of a free bottom quark plus corrections that are suppressed by inverse powers of the $b$ quark mass~\cite{Bigi:1992su,Bigi:1993fe,Bigi:1994wa,Manohar:1993qn}. 
The decay width of a $B$ meson with mass $M_B$ and four-momentum $p_B^\mu$
is given by
\begin{equation}
    \Gamma( B \to X_s \nu \bar \nu) = 
    \frac{1}{2M_B}
    \sum_X \int
    d\Phi_n \delta^{(4)}(p_B - p_X-p_\nu-p_{\bar\nu})
    \Big\vert  \langle X(p_X) \nu (p_\nu) \bar \nu (p_{\bar \nu})|
    \mathcal{H}_\mathrm{eff}
    | B(p_B)\rangle \Big\vert^2,
\end{equation}
where $\mathcal{H}_\mathrm{eff}$ is the effective Hamiltonian from Eq.~\eqref{eq:EFTdef}, $d\Phi_n$ denotes the phase-space integration and we
have summed over all possible final states $X$ into which the $B$ meson can decay.
Via the optical theorem and within the HQE framework one obtains 
\begin{align}
    \Gamma(B\to X_s \nu \bar{\nu}) &= 
    \Gamma_3
    +\Gamma_5 \frac{\langle O_5 \rangle}{m_b^2}
    +\Gamma_6 \frac{\langle O_6 \rangle}{m_b^3}
    +16 \pi^2 \tilde \Gamma_6 \frac{\langle \tilde O_6 \rangle}{m_b^3}
    +\dots
    \label{eq:decayrate1}
\end{align}
The HQE matrix elements are defined by $\braket{O_i} =\langle{B}(v)|O_i |{B}(v)\rangle/(2M_B)$, the forward matrix elements of the $\Delta B = 0$ operators between QCD states with momentum $p_B=M_B v$. 
The coefficients $\Gamma_i$ and $\tilde \Gamma_i$ describe the corresponding 
short distance contributions and can be calculated in series of $\alpha_s$.
We use the phase redefinition $b(x) = e^{-i m_b v \cdot x} b_v(x)$ to remove the large fraction of the $b$-field momentum.
The leading term $\Gamma_3$ describes the partonic $b \to s \nu \bar \nu$ decay
as  $\langle \bar{b}_v b_v \rangle=1$ is normalized to all orders in $\alpha_s$ and to lowest order in $1/m_b$. Up to order $1/m_b^3$, there are four operators 
independent on the spectator quark. The matrix elements are defined by
\begin{align}
\mu_\pi^2 &= -\langle \, \bar{b}_v (i D_\perp)^2 b_v \, \rangle \, ,  \\
\mu_G^2 &= \langle \, \bar{b}_v (-i S_{\mu\nu}) [iD_\mu, iD_\nu]\, b_v \, \rangle \, ,  \\
\rho_D^3 &= \langle \, \bar{b}_v \, [ i D_\mu^\perp , [ i v \cdot D, i D_\perp^\mu ]]\, b_v \, \rangle / 2,  \\
\rho_{LS}^3 &= \langle \, \bar{b}_v (-iS_{\mu \nu}) \{ iD^\mu_\perp,  [iv\cdot D,D_\perp^\nu]  \} \,  b_v \, \rangle \, ,
\end{align}
where $D_\mu^\perp=D_\mu-(v\cdot D)v_\mu$ and $S_{\mu \nu}=\sigma_{\mu \nu}/2=(i/4)[\gamma_\mu,\gamma_\nu]$. 

In heavy-to-light decay, starting from $1/m_b^3$ the HQE contains also four-quark operators $\tilde O_6$ which are phase-space enhanced (as indicated by the factor $16 \pi^2$). We have the following $\Delta B = 0$ operators:
\begin{align}
    \tilde O_1^{(q)} &=
    \Big(\bar b_v \gamma^\mu (1-\gamma_5) q\Big) 
    \Big(\bar q \gamma_\mu (1-\gamma_5) b_v\Big), \\
    \tilde O_2^{(q)} &=
    \Big(\bar b_v \slashed v (1-\gamma_5) q\Big)
    \Big(\bar q  \slashed v (1-\gamma_5) b_v\Big),  \\
    \tilde O_3^{(q)} &=
    \Big(\bar b_v^\alpha \gamma^\mu (1-\gamma_5) q^\beta\Big)
    \Big(\bar q^\beta \gamma_\mu (1-\gamma_5) b_v^\alpha\Big), \\
    \tilde O_4^{(q)} &=
    \Big(\bar b_v^\alpha \slashed v (1-\gamma_5) q^\beta\Big)
    \Big(\bar q^\beta  \slashed v (1-\gamma_5) b_v^\alpha\Big),
    \label{eqn:4qOperators}
\end{align}
where $\alpha, \beta $ are color indices. The matrix elements of four-quark 
operators can be parameterized in the following way:\footnote{This parametrization actually refers to the operators strictly defined in the $m_b \to \infty $ limit (HQET), while the operators in Eq.~\eqref{eqn:4qOperators} are  defined in QCD. Since in the following $\tilde \Gamma_6$ is considered only at LO and we neglect higher $1/m_b$ terms, we can adopt this approximation.}
\begin{align}
    \langle B_q | \tilde O_i^{q} |B_q  \rangle &= F^2_q(\mu_0) M_{B_q} \tilde B_i^{q}(\mu_0), \notag \\
    \langle B_q | \tilde O_i^{q'} |B_q  \rangle &= F^2_q(\mu_0) M_{B_q} \tilde \delta_i^{q'q}(\mu_0), \quad \text{for } q' \neq q,
\end{align}
where $\tilde B_i^{q}(\mu_0)$ and $\tilde \delta_i^{q'q}(\mu_0)$ the corresponding Bag parameters and ``eye-contractions'' evaluated at $\mu_0$, the renormalisation scale of the $\Delta B = 0$ operators, and $F_q(\mu_0)$ is the HQET decay constant:
\begin{equation}
    f_B = \frac{F_q(\mu_0)}{\sqrt{M_{B_q}}}
    \left[ 
    1+ \frac{\alpha_s}{2\pi}\left(
    \log \frac{m_b^2}{\mu_0^2} -\frac{4}{3}
    \right)
    +O\left( \frac{1}{m_b}\right)
    \right] 
    .
\end{equation}
For the decay $B \to X_s \nu \bar \nu$ we consider $B^0$ or $B^\pm$ mesons, with valance quark $q=u,d$ while the light quark in the four-quark operator is $q'=s$. In our analysis we use the estimates obtained from HQET sum rules~\cite{King:2021jsq} (see also \cite{Kirk:2017juj,Egner:2024lay,Black:2024bus}):
\begin{align}
    \tilde \delta^{su}_{1} &= \tilde \delta^{sd}_{1} = 0.0023^{+0.0140}_{-0.0091},
    &
    \tilde \delta^{su}_{2} &= \tilde \delta^{sd}_{2} = -0.0017^{+0.046}_{-0.0070},
\end{align}
where the values are given at the scale $\mu_0=1.5$~GeV. The LO contribution of four-quark operators can be summarized by the parameter
\begin{equation}
    \tau_0 = 8 \rho_D \log \left( \frac{\mu_0^2}{m_b^2} \right)
    + 32 \pi^2 \left( \langle \tilde O_2^s \rangle 
    - \langle \tilde O_1^s \rangle\right)
    =
    8 \rho_D \log \left( \frac{\mu_0^2}{m_b^2} \right)
    +16 \pi^2 F_B^2(\mu_0) \Big(\tilde \delta_2 - \tilde \delta_1 \Big)
    .
\end{equation}
The scale $\mu_0 \sim 1$~GeV is the $\overline{\mathrm{MS}}$
renormalization scale associated to the mixing between $\rho_D$ and the four-quark operators. 
The $\mu_0$ dependence cancels against the implicit scale dependence of $\langle \tilde O_1^s \rangle$ and $\langle \tilde O_2^s \rangle$ at $O(\alpha_s^0)$~\cite{Gambino:2005tp,Gambino:2007rp}.
The combination defined by $\tau_0$ is invariant under LO renormalization~\cite{Fael:2019umf} and 
does not depend on $\tilde O_3^{(q)}$ and $\tilde O_4^{(q)}$.
We obtain
\begin{equation}
    \tau_0 = -3^{+1.4}_{-0.6} \, \mathrm{GeV}^3 .
\end{equation}
where the uncertainty is dominated by of the bag parameters $\tilde \delta_i$
and we assumed that the uncertainties of $\tilde \delta_1$ and $\tilde \delta_2$ are
uncorrelated.

\subsection{Total rate}
\label{sec:TotalRate}

Approximating $m_s=0$ and neglecting QCD corrections to the power corrections, the total rate is given by
\begin{multline}
    \Gamma(B\to X_s \nu \bar{\nu}) = \Gamma_0 \left\{ 1 + C_F\sum_{n=1} X_n \left(\frac{\alpha_s}{\pi}\right)^n 
    - \frac{\mu_\pi^2}{2m_b^2} 
    - \frac{3\mu_G^2}{2m_b^2} 
    + \frac{3\rho_{LS}^3}{2m_b^3} 
    +  \frac{77}{6}
    \frac{\rho_D^3}{m_b^3}
    +\frac{\tau_0}{m_b^3}
    \right\} \label{eq:decayrate}
\end{multline}
in terms of tree level decay rate expression
\begin{align}
\Gamma_0 = N_\nu \frac{G_F^2 m_b^5}{192\pi^3} |V_{ts}V_{tb}|^2 |C_\nu|^2 \, ,
\end{align}
where $N_\nu=3$ is the number of neutrino species and $C_\nu$ is the Wilson coefficient of the Hamiltonian defined in Eq.~\eqref{eq:EFTdef}. The QCD corrections to the decay of the free quark up to N3LO~\cite{vanRitbergen:1999gs,Pak:2008cp,Fael:2024wxv,Chen:2023dsi} are described by the coefficients $X_1$, $X_2$ and $X_3$, with the color factor $C_F=4/3$,
and $\alpha_s \equiv \alpha_s^{(5)}(\mu_s)$ is the strong coupling
constant evaluated at the renormalization scale $\mu_s$.

The coefficients of the perturbative corrections $X_n$ for $n\geq 2$ introduce a dependence on the charm quark mass by the insertion of charm loops into internal gluon lines, expressed through the heavy quark mass ratio $\rho_c=m_c/m_b$. At NNLO, the expression, the so-called $U_C$ contribution, is known analytically in the pole scheme~\cite{Pak:2008cp}. For three light flavors ($n_l=3$), one massive charm ($n_c=1$) and one bottom quark ($n_h=1$), the perturbative corrections are (at the scale $\mu_s = m_b$)~\cite{vanRitbergen:1999gs,Pak:2008cp}
\begin{align}
X_1 &= \frac{25}{8} - \frac{\pi^2}{2} \simeq -1.8098 \, , 
\label{eqn:X1}
\\
X_2 &= 
\frac{933143}{15552}
-\frac{12403 \pi ^2}{1944}
-\frac{6671 \zeta (3)}{216}
-\frac{53}{36} \pi ^2 \log (2)
+\frac{289 \pi^4}{864}
\notag \\ & \quad 
+n_c
\left(-\frac{1009}{576}
+\frac{77 \pi ^2}{432}
+\frac{4 \zeta (3)}{3}
+\Delta X_2(\rho_c)
\right) 
\notag \\ & \simeq
-17.5818 + n_c(1.61017 +\Delta X_2(\rho_c)).
\label{eqn:X2}
\end{align}
The term denoted by $n_c$ arises from diagrams  with the insertion of a charm quark into a gluon propagator. We keep it separated for further discussion about
the charm quark effects in the total rate.
The function that describes the departure of the massless limit for the charm is defined such that $\Delta X_2(0)=0$.
As a reference value, for $\rho_c=0.218$ (see Tab.~\ref{tab:inputs}) we have 
$\Delta X_2=-0.794483$ and $X_2 = -17.5818 + 0.8166 n_c$.
Compared to the rest of $X_2$, the charm effect is rather small. Nevertheless we keep such effects for consistency since we will later convert the expression in the pole scheme to the kinetic scheme using the dependence on the charm mass at NNLO.

The third order correction is 
obtained by summing the exact fermionic contributions calculated in Ref.~\cite{Fael:2023tcv} and the analytic expression for the bosonic contribution 
in the large-$N_c$ limit from Ref.~\cite{Chen:2023dsi}. 
Moreover, we add the subleading color terms which result from the calculation in
Ref.~\cite{Fael:2020tow} for the $b\to cl\bar \nu_l$ decay rate expressed as a 
series expansion in $1-m_c^2/m_b^2\ll1$ and taking the limit $m_c = 0$.
For $n_l=3$ massless quarks we obtain
\begin{align}
    X_3&=-248.0 \, (2.0).
\end{align}
The quoted uncertainty arises from the massless extrapolation
of the subleading $N_c$ terms (see the discussion in~\cite{Fael:2023tcv}). The charm-mass effect to order $\alpha_s^3$ is currently unknown.

Finally, let us consider the approximations leading to Eq.~\eqref{eq:decayrate}, namely that we take $m_s=0$ and neglect electromagnetic corrections to the matrix element. 
The strange mass appears at leading order and the effect is to rescale the leading order term by about $1-8m_s^2/m_b^2 \simeq 0.9965$. The long-distance QED correction here is finite (unlike in $b \to u \ell \nu)$ and enters in the same way as the NLO QCD correction via a term proportional to $X_1$ but suppressed by the small value of $\alpha$ and the product of the quark charges $Q_b Q_s =1/9$. The QED correction is approximately $0.5\%$ of the QCD one. In summary, both the effects from the strange quark mass and long-distance QED are below the percent level and have been neglected.

\subsection{Kinetic scheme}
\label{sec:kinetic}
We express the total rate in terms of the kinetic bottom quark mass $m_b^\mathrm{kin}$ and the $\overline{\mathrm{MS}}$ mass for the charm.
The relation between the pole mass and the kinetic mass is given by
\begin{equation}
    m_b^\mathrm{pole} = 
    m_b^\mathrm{kin}(\mu) + [\overline \Lambda (\mu)]_\mathrm{pert}
    +\frac{[\mu_\pi^2(\mu)]_\mathrm{pert}}{2 m_b^\mathrm{kin}(\mu)}
    +O\left( \frac{1}{m_b^2} \right)\ ,
        \label{eqn:mpole2mkin}
\end{equation}
where the Wilsonian cutoff $\mu$ is a scale chosen such that 
$\Lambda_\mathrm{QCD} \ll \mu \ll m_b$.
The last two terms in \eqref{eqn:mpole2mkin} labeled by ``pert'' are 
the HQE parameters calculated in perturbative QCD with the Small Velocity sum rules~\cite{Bigi:1994ga}. 
The analytic expressions up to $O(\alpha_s^3)$ are~\cite{Czarnecki:1997sz,Fael:2020iea,Fael:2020njb}:
\begin{align}
[\overline \Lambda (\mu)]_\mathrm{pert} &=
\mu \left[\frac{16}{9} \frac{\alpha_s}{\pi}
+\left(\frac{\alpha_s}{\pi}\right)^2 \left(\frac{244}{9}-\frac{8 \pi ^2}{9}-8 \lmu-\frac{8 }{27}\lc n_c\right)
+\left(\frac{\alpha_s}{\pi}\right)^3 \left(
\frac{10373}{18}
\right. \right. 
\notag \\[5pt] & \quad
-\frac{274 \pi ^2}{9}
-\frac{886 \zeta_3}{9}
+\frac{2 \pi ^4}{3}
-\frac{2324}{9} \lmu
+8 \pi ^2  \lmu
+36 \lmu^2
+\frac{4 }{81}\lc^2 n_c^2
\notag \\[5pt] & \quad
\left. \left.
+n_c \left(-\frac{14}{27}-\frac{94}{9} \lc
+\frac{8 }{3}\lc \lmu
+\frac{8 \pi ^2}{27} \lc \right)
\right)\right],
\label{eqn:lambdaBarpert}
\\
[\mu_\pi^2(\mu)]_\mathrm{pert} &=
\mu^2 \left[\frac{4 }{3}\frac{\alpha_s}{\pi}
+\left(\frac{\alpha_s}{\pi}\right)^2 
\left(\frac{52}{3}-\frac{2 \pi ^2}{3}-6 \lmu-\frac{2 }{9}\lc n_c\right)
+\left(\frac{\alpha_s}{\pi}\right)^3 
\left(\frac{2575}{8}
\right. \right.
\notag \\[5pt] & \quad 
-\frac{119 \pi ^2}{6}
-\frac{443 \zeta_3}{6}
+\frac{\pi ^4}{2}
-\frac{500 }{3}\lmu
+6 \pi ^2\lmu 
+27 \lmu^2
+\frac{1}{27}\lc^2 n_c^2
\notag \\[5pt] & \quad 
\left. \left.
+n_c \left(-\frac{7}{18}-\frac{41 }{6}\lc+2 \lc \lmu
+\frac{2 \pi ^2}{9} \lc
\right)\right)\right],
\label{eqn:mupipert}
\\
[\rho_D^3(\mu)]_\mathrm{pert} &=
\mu^3 \left[\frac{8 }{9}\frac{\alpha_s}{\pi}
+\left(\frac{\alpha_s}{\pi}\right)^2 
\left(\frac{98}{9}
-\frac{4 \pi ^2}{9}
-4 \lmu-\frac{4 }{27}\lc n_c
\right)
+\left(\frac{\alpha_s}{\pi}\right)^3 \left(
\frac{20959}{108}
\right. \right.
\notag \\[5pt] & \quad 
-\frac{113 \pi ^2}{9}
-\frac{443 \zeta_3}{9}
+\frac{\pi ^4}{3}
-\frac{946 }{9}\lmu
+4  \pi ^2\lmu
+18 \lmu^2
+\frac{2 }{81}\lc^2 n_c^2
\notag \\[5pt] & \quad 
\left.
+n_c \left(-\frac{7}{27}-\frac{13 \lc}{3}+\frac{4 \lc \lmu}{3}+\frac{4 \lc \pi ^2}{27}\right)
\right)\Bigg],
\label{eqn:rhoDpert}
\end{align}
where here $\alpha_s=\alpha_s^{(4)}(\mu_s)$, $l_\mu = \log(2 \mu/\mu_s)$ and $l_{m_c} = \log(\mu_s^2/(m_c^\mathrm{pole})^2)$. In our implementation, we follow scheme B of Ref.~\cite{Fael:2020njb} 
and include also the decoupling effects of the charm quark in the kinetic scheme
(denoted by $n_c=1$).
In the kinetic scheme, also the HQE parameters are redefined in the following way:
\begin{align}\label{eq:rhodpert}
    \mu_\pi^2(0) &= \mu_\pi^2(\mu) - [\mu_\pi^2(\mu)]_\mathrm{pert}\ , &
    \rho_D^3(0) &= \rho_D^3(\mu) - [\rho_D^3(\mu)]_\mathrm{pert}\ .
\end{align}
For the charm mass, we need the relation between the pole and $\overline{\mathrm{MS}}$ up to $O(\alpha_s)$:
\begin{equation}
m_c^\mathrm{pole} = 
\overline{m}_c(\mu_c) 
\left[ 
1 
+ \frac{\alpha_s^{(4)}(\mu_s)}{\pi}
\left(\frac{4}{3} + \log\left( \frac{\mu_c^2}{\overline{m}_c^2(\mu_c)} \right) \right) \right].
\label{eqn:mpole2mMS}
\end{equation}
We replace $m_b^\mathrm{pole}$ and $m_c^\mathrm{pole}$
in the total rate formula using~\eqref{eqn:mpole2mkin} 
and~\eqref{eqn:mpole2mMS} and re-expand it as a series in $\alpha_s^{(4)}$.

After the scheme conversion, the charm mass effect at NNLO receives a contribution
from the genuine $n_c$ appearing in $X_2$ in Eq.~\eqref{eqn:X2}
as well as the terms from the conversion of the bottom mass in Eqs.~(\ref{eqn:lambdaBarpert}-\ref{eqn:rhoDpert}). 
To order $\alpha_s^3$ we can in principle calculate the $n_c$ 
terms arising from the mass conversion.
However, since the genuine $n_c^2$ and $n_c$ contributions to $X_3$ are unknown, 
we consistently neglect all charm-mass effects at $O(\alpha_s^3)$, 
including those induced by scheme conversion.

As a reference, we obtain, 
with $m_b^\mathrm{kin} (1\, \mathrm{GeV}) = 4.576$~Gev, $\overline{m}_c(2\, \mathrm{GeV}) = 1.090$~GeV and $\mu_s=m_b^\mathrm{kin}/2$,
the following prediction for the matrix elements:
\begin{equation}
\Gamma(B \to X_s \nu \bar \nu)/\Gamma_0 =
1 - 0.0360_{\alpha_s}
+ (0.0216 - 0.00020 n_c)_{\alpha_s^2}
+ 0.0237_{\alpha_s^3} 
- 0.0097_{pw}.
\label{eqn:GammaNumericalExample}
\end{equation}
The charm quark contribution at $O(\alpha_s^2)$ is denoted by $n_c$, while the last term
denoted by \textit{pw} is the overall contribution from power corrections (with $\tau_0 = 0$). From the numerical result in Eq.~\eqref{eqn:GammaNumericalExample}
we observe that the overall contribution from charm quarks in the loop is rather small at $O(\alpha_s^2)$. We verified that this observation does not depend on our choice for the quark masses and scale of $\alpha_s$. Instead, it indicates that the charm quark, when is treated consistently in both the total rate and the short-distance mass definition, effectively tends to decouple from the problem.  In retrospect, this justifies our choice of $n_\ell=3$ in the $O(\alpha_s^3)$ corrections. The neglected $n_c$ contribution at third order is likely to be rather small.

\begin{figure}[htb]
    \centering
    \includegraphics[width=1\linewidth]{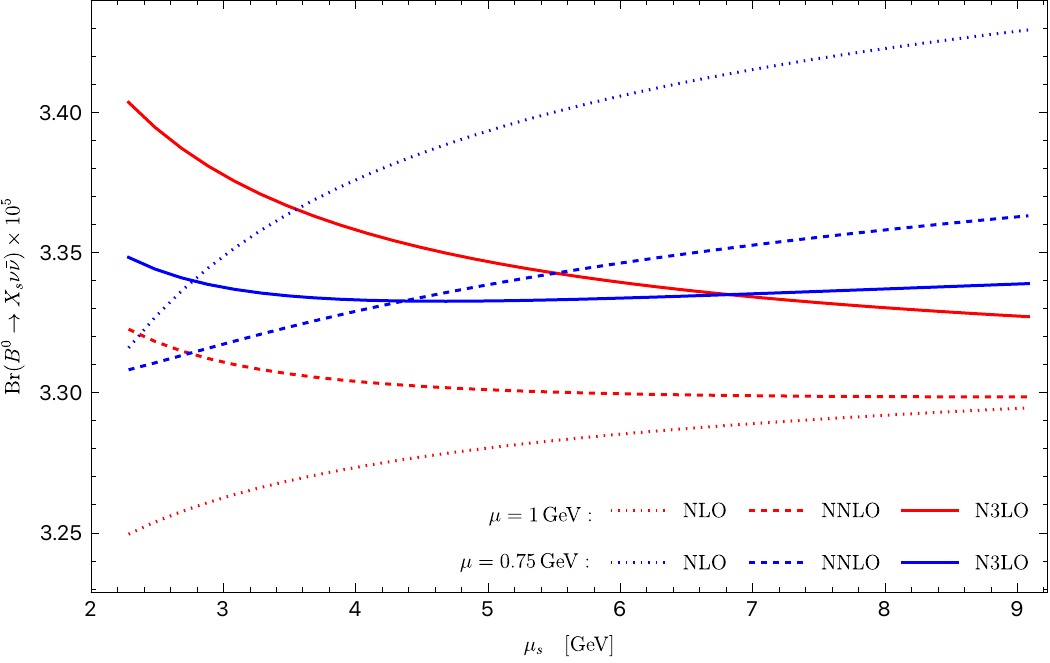}
    \caption{The dependence of the branching ratio $\mathrm{Br}(B^0 \to X_s \nu \bar \nu)$ on the renormalization scale $\mu_s \simeq m_b$ at the low scale
    (the scale of $C_\nu$ is kept fixed at $M_Z$).
    In the plot, we compare two predictions at different values of the Wilsonian cutoff $\mu$
    in the kinetic scheme: $\mu=1$~GeV and $\mu=0.75$~GeV. }
    \label{fig:plotmus}
\end{figure}

In Fig.~\ref{fig:plotmus} we show the dependence of the branching ratio on the renormalization scale $\mu_s$, varied in the range $\mu_s \in [m_b^{\mathrm{kin}}/2,, 2 m_b^{\mathrm{kin}}]$, while keeping the scale of $C_\nu$ fixed at $M_Z$. Results are shown at different orders in the perturbative expansion.
The red curves correspond to our NLO, NNLO, and N3LO predictions obtained using a Wilsonian cutoff $\mu = 1~\mathrm{GeV}$ in the kinetic scheme. This is the default choice adopted in the global fits of Refs.~\cite{Bernlochner:2022ucr,Finauri:2023kte} to extract $m_b^{\mathrm{kin}}(\mu)$ and the HQE parameters. At NLO (NNLO) we observe a $0.7\%$ ($0.4\%$) variation with respect to the central value at $\mu_s = m_b^{\mathrm{kin}}$, while at N3LO the variation increases to $1.1\%$, exceeding the NLO one.

A more stable perturbative behavior is obtained if the kinetic mass and the HQE parameters are renormalized at the lower scale $\mu = 0.75~\mathrm{GeV}$. The corresponding results are shown by the blue curves in Fig.~\ref{fig:plotmus}. To this end, we recompute $m_b^{\mathrm{kin}}(0.75~\mathrm{GeV})$ using \texttt{RunDec}, and evolve $\mu_\pi^2$ and $\rho_D^3$ from the values listed in Tab.~\ref{tab:inputs}, exploiting the fact that the left-hand side of Eq.~\eqref{eqn:rhoDpert} is $\mu$-independent:
\begin{align}
    \mu_\pi^2(\mu_1)
    =\mu_\pi^2(\mu_2) + [\mu_\pi^2(\mu_1)]_\mathrm{pert}- [\mu_\pi^2(\mu_2)]_\mathrm{pert},
    \notag\\
    \rho_D^3(\mu) 
    =\rho_D^3(\mu_2) + [\rho_D^3(\mu_1)]_\mathrm{pert} - [\rho_D^3(\mu_2)]_\mathrm{pert}.
\end{align}
With this choice, the scale variation is reduced to $1.7\%$, $0.8\%$, and $0.3\%$ at NLO, NNLO, and N3LO, respectively.

Nevertheless, a scale of $\mu = 0.75~\mathrm{GeV}$ lies at the boundary of the perturbative regime, where $\Lambda_{\mathrm{QCD}} \ll \mu$ is only marginally satisfied, and is therefore difficult to justify as a default choice in our setup. Despite this, the near-perfect overlap of the N3LO curves for $\mu = 1~\mathrm{GeV}$ and $\mu = 0.75~\mathrm{GeV}$ (red and blue solid lines in Fig.~\ref{fig:plotmus}) indicates that a consistent prediction for the branching ratio is achieved once $\mathcal{O}(\alpha_s^3)$ corrections are included.
We thus conclude that the choice $\mu = 1~\mathrm{GeV}$ provides a satisfactory prediction, albeit with a likely conservative estimate of the perturbative uncertainty from scale variation. Moreover, since the fit of Ref.~\cite{Finauri:2023kte} was performed at $\mu = 1~\mathrm{GeV}$, we adopt this value to consistently use the quoted uncertainties and correlations among the HQE parameters without further modifications.

\subsection{The spectrum}
\label{sec:spectrum}

The spectrum of semileptonic $b \to u l \bar \nu_l$ decays is currently known at NNLO. The differential decay rate, normalized to the tree level expression for the total rate is naively given in the pole scheme in terms of distribution-valued functions of $s=q^2/m_b^2$.
Although the $q^2$ distribution is divergent at the endpoint $q^2 \sim m_b^2$, the branching ratio with a lower cut on the neutrino invariant mass can be calculated in the HQE as
\begin{align}
\Gamma(q^2>q_0^2)&=\int_{q_0^2}^{M_B^2} dq^2\frac{d\Gamma}{dq^2} = \int_{s_0}^{\infty} ds \,\frac{d\Gamma_{\textrm{OPE}}}{ds} \, ,
\label{eqn:GammaCut}
\end{align}
where $s_0=q_0^2/m_b^2$ and we have neglected QCD corrections to the power corrections. The upper limit of the integral on the RHS is effectively cut off at around $s \sim 1$, but Eq.~\eqref{eqn:GammaCut} is technically correct as written due to the distributions in $d\Gamma_{\textrm{OPE}}/ds$ which have support on a finite interval. The RHS of Eq.~\eqref{eqn:GammaCut} can then be calculated as
\begin{align}
    &\Gamma(q^2>q_0^2)=
    \Gamma_0 \Bigg\{ X_0(s_0) \left(1- \frac{\mu_\pi^2}{2m_b^2}\right)
    +C_F \sum_{n\geq 1}X_n(s_0) \left( \frac{\alpha_s}{\pi} \right)^n  \nonumber \\
    & 
    + X_G(s_0) \frac{\mu_G^2}{m_b^2} + X_{LS}(s_0) \frac{\rho_{LS}^3}{m_b^3} + X_D(s_0) \frac{\rho_D^3}{m_b^3} + \frac{\tau_0}{m_b^3} \Bigg\}
\end{align}
involving regular functions $X_i(s_0)$.
At tree level, we have~\cite{Fael:2019umf}
\begin{align}
    X_0(s_0) &= (1-s_0)^3(1+s_0) \, , \\
    X_G(s_0) &= -X_{LS}(s_0) =  \frac{1}{2}(1 - s_0) (-3 - 5 s_0 - 5 s^2_0 + 5 s^3_0),\\
    X_{D}(s_0) &=
    \frac{77}{6}
    -\frac{37 s_0}{3}
    -4s_0^2
    -\frac{11 s_0^3}{3}
    -\frac{5 s^4_0}{6}
    -16 \log (1-s_0).
\end{align}
The NLO coefficient function is
\begin{align}
    X_1(s_0) &=\frac{25}{6}-\frac{61 s_0}{9}-\frac{s_0^2}{18}+\frac{11 s_0^3}{3}-s_0^4
    +\pi ^2 \left(-\frac{2}{3}+\frac{8 s_0}{9}-\frac{8 s_0^3}{9}+\frac{4 s_0^4}{9}\right)
    \nonumber \\ & \quad
    +\left(-\frac{4 s_0}{3}+2 s_0^2-\frac{4 s_0^3}{9}-\frac{5 s_0^4}{3}\right) \log (s_0)
    +\left(-\frac{4}{3}+\frac{16 s_0}{3}-\frac{16 s_0^3}{3}+\frac{8 s_0^4}{3}\right) \text{Li}_2(s_0)
    \nonumber \\ & \quad
    + \left(-\frac{43}{9}+\frac{28 s_0}{3}-4 s_0^2-\frac{20 s_0^3}{9}+\frac{5 s_0^4}{3}\right) \log (1-s_0)
    \nonumber \\ & \quad 
    +\left(-\frac{4}{3}+\frac{8 s_0}{3}-\frac{8 s_0^3}{3}+\frac{4 s_0^4}{3}\right) \log (s_0)\log (1-s_0),
\end{align}
which reduces to $X_1(0)=X_1$ given in Eq.~\eqref{eqn:X1} when the cut is released ($s_0=0$). The expression for $X_2$ can be calculated from the analytical form of the differential rate given in terms of HPLs in Ref.~\cite{Chen:2022wit} and in turn can be expressed in terms of HPLs. We do not give the explicit expression here as it is quite involved. As far as the charm mass effect which appears at NNLO, we simply set $n_c=0$ in $X_2$ and assume that the effect is under control (c.f. Eq.~\eqref{eqn:GammaNumericalExample}). 

\section{Results}
\label{sec:results}
The inputs used in our analysis are summarized in Table~\ref{tab:inputs}. For the strong coupling, we prefer to use the PDG average of $\alpha_s(M_Z)$, although we note that a very precise determination from lattice QCD has been reported~\cite{Brida:2025gii}. The CKM factor $|V_{ts}^* V_{tb}|$ can be written in terms of $|V_{cb}|$ and the ratio $|V_{ts}^* V_{tb}|/|V_{cb}|=0.965(1)$ which is precisely calculated (assuming CKM unitarity) using the Wolfenstein expansion $|V_{ts}^* V_{tb}|/|V_{cb}|=1+\lambda^2(2\bar{\rho}-1) +O(\lambda^4)$ taking $\lambda=0.22519$ and $\bar{\rho}=0.1609(95)$ from UTFit~\cite{UTfit:2022hsi} and neglecting the $O(\lambda^4)$ terms and the uncertainty on $\lambda$. 

\begin{table}[ht!]
\centering
\begin{tabular}{@{}ll ll@{}}
\toprule
$\alpha_s(M_Z)$ & $0.1180(9)$ & $|V_{cb}|$ & $0.04197(48)$~\cite{Finauri:2023kte} \\
$\alpha(M_Z)$ & $1/127.952$~\cite{Davier:2019can}& $|V_{ts}^* V_{tb} |/ |V_{cb}|$ & $0.965(1)$~\cite{UTfit:2022hsi}  \\
$\sin^2\theta_W(M_Z)$ & $0.23141(4)$ & $\tau_{B^+}$ & $1.638(4) \, \ps$ \\
$M_H$ & $125.20 \, \GeV$ & $\tau_{B^0}$ & $1.517(4) \, \ps$ \\
$G_F$ & $1.166 378 8 \times 10^{-5} \,\GeV^{-2}$ & $m_b^{\text{kin}}$ & $4.573(12) \, \GeV$~\cite{Finauri:2023kte} \\
$M_Z$ & $91.1880 \, \GeV$ & $\mu_\pi^2$ & $0.454(43) \, \GeV^2$~\cite{Finauri:2023kte} \\
$M_W$ & $80.3692 \, \GeV$ & $\mu_G^2$ & $0.288(49) \, \GeV^2$~\cite{Finauri:2023kte}  \\
$\overline{m}_t(\overline{m}_t)$ & $162.87(60)\, \GeV$ & $\rho_D^3$ & $0.176 (19)\, \GeV^3$~\cite{Finauri:2023kte}  \\
$\overline{m}_b(\overline{m}_b)$ & $4.200(14) \, \GeV$~\cite{FlavourLatticeAveragingGroupFLAG:2024oxs} & $\rho_{LS}^3$ & $ -0.113(90) \, \GeV^3$~\cite{Finauri:2023kte}    \\
$m_b/m_c$ & 4.579(9)~\cite{FlavourLatticeAveragingGroupFLAG:2024oxs}&  $\tilde\delta^{sd}_2-\tilde\delta^{sd}_1$ & $-0.0040(160)$ \, \cite{King:2021jsq} \\
\bottomrule
\end{tabular}
\caption{Table of inputs used in the numerical analysis. The gauge couplings and quark masses are renormalized in the five-flavor theory, with the exception of the top mass and the mixing angle which are defined for six active flavors.  The HQE parameters are taken from a fit to the normalized kinematical distributions of $B \to X_c \ell \nu$ and are correlated with $|V_{cb}|$ and $m_b^{\text{kin}}$ -- see Table 4 of Ref.~\cite{Finauri:2023kte} for full correlation matrix. Unless otherwise specified the inputs are taken from the PDG~\cite{ParticleDataGroup:2024cfk} aside from the $\overline{\text{MS}}$ value of $\overline{m}_t(\overline{m}_t)$ (see text for details). Values without quoted uncertainties are not varied in our uncertainty analysis.
}
\label{tab:inputs}
\end{table}

Using the expression for the total rate in Eq.~\eqref{eq:decayrate} and the inputs in Table~\ref{tab:inputs}, we obtain the inclusive branching ratios,
\begin{align}
    \mathrm{Br}(B^0 \to X_s \nu \bar \nu)\times 10^5 &= 3.351\pm 0.058_{V_{cb}}\pm 0.029_{\text{param}}\pm 0.043_{\text{HQE}}\pm 0.065_{\text{scale}}  \, ,\label{eq:brError}
\end{align}
The quoted errors arise respectively from $V_{cb}$ (including its correlation with other HQE parameters; discussed below), the top mass $\overline{m}_t(\overline{m}_t)$, the HQE matrix elements and kinetic mass fitted from moments of semileptonic $B \to X_c \ell \nu$ decay in Ref.~\cite{Finauri:2023kte}, and the uncertainty from varying the electroweak matching scale $\mu_{\text{EW}}\in[60\text{ GeV},320\text{ GeV}]$ ($\delta_{\mu_\text{EW}}=0.045$), the scale at which $\alpha_s$ is evaluated $\mu_s\in[m_{b,\text{kin}}(1\text{ GeV})/2,2m_{b,\text{kin}}(1\text{ GeV})]$ ($\delta_{\mu_s} = 0.040$), and the scale at which $m_{b,kin}$ is evaluated $\mu_{\text{kin}}\in[0.8\text{ GeV}, 1.2\text{ GeV}]$ ($\delta_{\mu_{\text{kin}}} = 0.028$). Errors arising from other parametric inputs are negligible ($\delta_{\text{param.}}< 0.01$) so we do not include them in Eq.~\eqref{eq:brError}. The corresponding branching ratio for the charged-mode decay can be trivially obtained from Eq.~\eqref{eq:brError} using the ratio $\tau_{B^+}/\tau_{B^0}$ (the lifetime uncertainties are negligible).

Combining these sources of error gives a $\sim 3\%$ uncertainty in the final SM prediction of the branching ratios of the $B^0$ and $B^+$ mesons,
\begin{align}
    \mathrm{Br}(B^0 \to X_s \nu \bar \nu) &= (3.351\pm 0.102)\times 10^{-5}\, , \label{eq:totalBr}\\
    \Br (B^+ \to X_s \nu \bar{\nu}) &= (3.618 \pm 0.110) \times 10^{-5} \, . \label{eq:totalBr+}
\end{align}
The partial rates are equal up to isospin breaking corrections to forward matrix elements in the HQE. We decline to make a prediction for the branching ratio $\Br(B_s \to X \nu \bar{\nu})$ sensitive to weak annihilation of the valence quarks $b\bar{s}\to \nu \bar{\nu}$ and $SU(3)$ breaking corrections to the HQE matrix elements. For the Cabibbo suppressed decay mediated by $b \to d \nu \bar \nu$ 
one can estimate the branching ratio as
$\mathrm{Br}(B \to X_d \nu \bar \nu) 
\simeq |V_{td}/V_{ts}|^2 \mathrm{Br}(B \to X_s \nu \bar \nu)$.
For the $B^+$, this relation is essentially exact up to $m_s^2$ corrections (c.f.\,Tables 5 and 7 of Ref.~\cite{King:2021jsq}) and 
one can simply rescale the determination in Eq.~\eqref{eq:totalBr+}
and take into account the uncertainty from $|V_{td}/V_{ts}|$.
Using the central value $|V_{td}/V_{ts}| = 0.2080$ from~\cite{UTfit:2022hsi}
we estimate $\mathrm{Br}(B^+ \to X_d \nu \bar \nu) = 1.56 \times 10^{-6}$.
For $B^0$, the $d$ is both a valence quark and a quark appearing in the four-quark operators, so the determination of $B^0 \to X_d \nu \bar \nu$ is not a simple 
rescaling of Eq.~\eqref{eq:totalBr}.

To obtain Eq.~\eqref{eq:totalBr}, we used the inclusive determination of $V_{cb}$ from Ref.~\cite{Finauri:2023kte}. However, since this value of $V_{cb}$ is known to be in tension from that obtained from exclusive-mode data, we also report the result of the branching ratio with a general dependence on $V_{cb}$ and its uncertainties
\begin{align}\label{eq:generalVcb}
    \mathrm{Br}(B^0 \to X_s \nu \bar \nu) &= 0.01902|V_{cb}|^2\pm |V_{cb}|\sqrt{\Delta_1\delta{V_{cb}}^2
    + \Delta_2\rho_{m_bV_{cb}}\delta{V_{cb}}|V_{cb}|+\Delta_3|V_{cb}|^2}
\end{align}
where
\begin{equation}
    \Delta_1 = 1.4477\times 10^{-3}\,,\quad
    \Delta_2 = 1.8997\times 10^{-5}\,,\quad
    \Delta_3 = 2.2586\times 10^{-7}\,,
\end{equation}
and $\delta V_{cb}$ and $\rho_{m_bV_{cb}}$ are the uncertainty of $|V_{cb}|$ and the correlation between $|V_{cb}|$ and the bottom-quark mass in the kinetic scheme, respectively. Note that there is also a correlation term between $|V_{cb}|$ and the other HQE parameters, but is about an order of magnitude smaller than the other sources of uncertainty, so we have neglected it in Eq.~\eqref{eq:generalVcb}.

With the same numerical setup as for the total rate, we obtain the following branching ratios with lower cuts $q^2>q_0^2$ (Table~\ref{tab:resultstable}). In addition we provide branching ratios with upper cuts which are obtained as the difference with respect to the total rate, and binned partial branching fractions, which are obtained as the difference between two lower cuts. The branching ratios away from the endpoint region $q^2 \sim m_b^2$ (where the $1/m_b$ expansion breaks down) are not sensitive to weak annihilation at this order. The differential branching fractions integrated over the endpoint region which we define as $q^2 >16\,\textrm{GeV}^2$ are, of course, sensitive to the $1/m_b$ corrections. Even away from the endpoint, the relative uncertainties of the branching fractions with an upper cut on $q^2$ are moderately larger than the total branching fractions, since the former indirectly probe the endpoint region through the difference with respect to the total rate.

\begin{table}[ht]
\centering
\begin{tabular}{l @{\hspace{8mm}} l @{\hspace{8mm}} l @{\hspace{8mm}} l}
$q^2$ bin & $B^+ \to X_s \nu \bar{\nu}$ & $B^0 \to X_s \nu \bar{\nu}$ & $B^+/B^0$ Avg. \\
\toprule
Total & $3.618 \pm 0.110$ & $3.351 \pm 0.102$ & $3.485 \pm 0.106$ \\
\midrule
$[>4]$ & $2.311\pm0.077$ & $2.141 \pm 0.071$ & $2.227 \pm 0.074$\\
$[>8]$ & $1.245 \pm 0.057$ & $1.153 \pm 0.053$ & $1.199 \pm 0.055$ \\
$[>12]$ & $0.484 \pm 0.043$ & $0.448 \pm 0.040$ & $0.466 \pm 0.041$\\
$[>16]$ & $0.066 \pm 0.040$ & $0.061 \pm 0.037$ & $0.064 \pm 0.039$\\
\midrule
$[<4]$ & $1.307 \pm 0.038$ & $1.210 \pm 0.036$ & $1.258 \pm 0.037$\\
$[<8]$ & $2.373 \pm 0.068$ & $2.198 \pm 0.063$ & $2.286\pm0.066$\\
$[<12]$ & $3.135 \pm 0.087$ & $2.903 \pm 0.081$ & $3.019 \pm 0.084$\\
$[<16]$ & $3.552 \pm 0.100$ & $3.290\pm 0.093$ & $3.421 \pm 0.097$\\
\midrule
$[4,8]$ & $1.067 \pm 0.035$ & $0.988 \pm 0.032$ & $1.027 \pm 0.034$\\
$[8,12]$ & $0.761 \pm 0.021$ & $0.705 \pm 0.020$ & $0.733 \pm 0.020$\\
$[12,16]$ & $0.417 \pm 0.035$ & $0.387 \pm 0.032$ & $0.402 \pm 0.033$\\
\bottomrule
\end{tabular}
\caption{
Summary of our final results for the total branching fractions of inclusive $B \to X_s \nu \bar{\nu}$ decays at $O(1/m_b^3,\alpha_s^3)$ and partial branching fractions in several bins of $q^2$ at $O(1/m_b^3, \alpha_s^2)$. The branching fractions are given in units of $10^{-5}$ and the bins in units of $\textrm{GeV}^2$. The averages are reference values assuming equal production of $B^+$ and $B^0$ mesons at the $\Upsilon(4S)$. In the Standard Model the CP asymmetries of these decays are negligible, so $B^+$ and $B^0$ refer to the charged and neutral meson decays in general. \label{tab:resultstable}}
\end{table}

Our results in Eqs.~\eqref{eq:totalBr} and~\eqref{eq:totalBr+} are significantly larger than the previous estimate of the branching ratio $\Br(B \to X_s \nu \bar{\nu}) = (2.9 \pm 0.3) \times 10^{-5}$ in the SM in Ref.~\cite{Buras:2014fpa} (see also the discussion in Ref.~\cite{Altmannshofer:2009ma}) and about three times more precise. Part of this difference can be attributed to parametric input; the authors of Ref.~\cite{Buras:2014fpa} used a smaller and more uncertain value of the CKM parameter $|V_{cb}|=0.0409(10)$ corresponding to an inflated average of inclusive and exclusive determinations available at the time. With this historical value of $|V_{cb}|$ (and dropping the correlation between $|V_{cb}|$ with HQE parameters including $m_b$) our prediction for the branching ratios with otherwise the same setup are decreased to $\Br(B^0 \to X_s \nu \bar{\nu}) = (3.182 \pm 0.177) \times 10^{-5}$ and $\Br(B^+ \to X_s \nu \bar{\nu}) = (3.436 \pm 0.191) \times 10^{-5}$ and the uncertainties increase.

The remaining difference could be attributed to an inconsistent 
use of the $1S$ scheme. In~\cite{Buras:2014fpa,Altmannshofer:2009ma} 
the branching ratio formula is obtained 
using the bottom quark mass in the $1S$ scheme (with $m_b^{1S}=4.66$~GeV) 
for the overall $m_b^5$ power in the rate,
and taking into account the NLO QCD corrections to the $b \to s \nu \bar \nu$ matrix element by multiplying the tree-level result by a factor $\kappa(0) = 1+C_F X_1 \alpha_s/\pi = 0.83$, strictly valid only in the pole scheme, instead of the appropriate factor in the 1S scheme: $\kappa^{1S}(0)= 1-0.115 = 0.885$~\cite{Hoang:1998ng}. Taking into account this additional 7\% upward shift, the branching ratio 
of~\cite{Buras:2014fpa} would change to $3.1 \times 10^{-5}$, in agreement
within uncertainties with our final results.

In addition, Ref.~\cite{Buras:2014fpa} does not include $1/m_b^3$ corrections 
which are positive and tends to compensate the negative shift from the perturbative QCD corrections. The values for the HQE parameters are taken from 
(in our opinion) a superseded fit of semileptonic $B$ decays from Ref.~\cite{Bauer:2004ve}.

\section{Conclusions}
\label{sec:conclusions}
In this paper we have presented a comprehensive study of the total rate and leptonic spectrum of the inclusive $B\to X_s \nu\bar\nu$ decay in the Standard Model. The underlying $b\to s \nu\bar\nu$ transition has recently garnered considerable interest because of a 2.7$\sigma$ anomaly observed in the exclusive $B\to K \nu\bar\nu$ mode~\cite{Belle-II:2023esi, Belle-II:2025lfq} which, besides offering a tantalizing potential connection to existing anomalies in the exclusive $B\to K^{(*)}\mu^+\mu^-$ modes~\cite{Capdevila:2023yhq, Hurth:2025vfx}, could also be the first evidence for potential dark matter~\cite{Berezhnoy:2025osn} or axion~\cite{Gao:2025ohi} candidates. The inclusive mode offers a test of the underlying physics which is completely independent both from the theoretical and experimental point of view. 

The updated theoretical prediction and the recent experimental upper limit~\cite{Belle-II:2025bho} are given in Eqs.~(\ref{eq:BRSMiso}) and (\ref{eq:BRexp}), respectively. We include NLO QCD and EW corrections to the relevant Wilson coefficient, N3LO corrections to matrix element (taken directly from existing $B\to X_u \ell\nu$ studies), and $1/m_b^2$ and $1/m_b^3$ power corrections.  

While the perturbative corrections to the Wilson coefficient are of the expected magnitude, the N3LO corrections to the matrix element turn out to be numerically significantly larger than what would be naively inferred from NNLO scale variations when the Wilsonian cutoff is chosen as $\mu = 1~\mathrm{GeV}$, as illustrated in Fig.~\ref{fig:plotmus}. We find that the convergence of the perturbative expansion up to N3LO improves when the cutoff $\mu$ is lowered. Whether this improvement is merely accidental or reflects a deeper physical mechanism remains to be clarified.
The well-controlled behavior of analogous higher-order corrections in the $b \to c \ell \nu$ decay rate at $\mu = 1~\mathrm{GeV}$ (see, for example, Ref.~\cite{Fael:2024rys}) suggests that the observed pattern may be linked to the definition of the kinetic $b$-quark mass. Indeed, the kinetic mass is tied to a universal function describing soft-gluon emission by heavy quarks in heavy-to-heavy transitions~\cite{Fael:2020iea,Fael:2020njb,Fael:2021yjc}. When applied to heavy-to-light processes such as $b \to u$ and $b \to s$ transitions, the kinetic scheme still ensures the cancellation of the leading renormalon, but it may lead to a slower convergence of the perturbative series. This can be attributed to the substantially different structure of real radiation in heavy-to-light decays, where gluons collinear to the massless final-state quark play an important role.
In this context, it is conceivable that a short-distance mass scheme specifically tailored to heavy-to-light transitions could reorganize the perturbative expansion and lead to a better-behaved series. Nevertheless, the numerical impact of these effects is at the level of a few percent, and we therefore expect them to be of greater phenomenological relevance for the extraction of $|V_{ub}|$ from inclusive semileptonic $B$ decays than for the present analysis of $B \to X_s \nu \bar{\nu}$.

Another issue of concern is that the $B\to X_s\nu\bar\nu$ rate can only be measured for $M_X < M_X^{\rm cut} \sim 2.0 \; {\rm GeV}$ and that the required extrapolation is currently based on a rather old one-parameter Fermi motion model. This situation is also present for $B\to X_u \ell\nu$ (whose spectrum is expected to be similar to the $B\to X_s \nu\bar\nu$ one) and $B\to X_s \ell^+\ell^-$ decays. In the literature, this problem has been addressed adopting two related but not identical strategies. In Ref.~\cite{Bernlochner:2020jlt}, the $B\to X_s \gamma$, $X_u\ell\nu$ and $X_s\ell^+\ell^-$ decays are described in terms of a universal leading power and several subleading shape functions. The latter contribute differently to each mode and need to be modeled. In Ref.~\cite{Gambino:2016fdy}, the authors adopt an approach based on a factorization ansatz in which each process depends on a single, albeit process and $q^2$ dependent, shape function. The difference between these shape functions is a subleading power effect and requires modeling. While both strategies are viable, it is not clear whether the required extrapolation for $B\to X_s (\ell\ell, \nu\bar\nu)$ modes can be performed without introducing this modern description of the $M_X$ spectrum into the EvtGen Monte Carlo~\cite{Lange:2001uf}, which is currently used to simulate these decays at Belle~II. 

Finally, we comment on future improvements of the experimental determination of the inclusive branching ratio at Belle II. Currently statistical and systematic uncertainties are, respectively, about a factor of three and four larger than the SM prediction. An order of magnitude luminosity increase (i.e.~5~ab$^{-1}$) should be quite achievable and would reduce the statistical uncertainty to about $2\times 10^{-5}$. The two largest sources of systematic and the simulated-sample size and background normalization contribute an uncertainty of about $6\times 10^{-5}$ and will benefit from more extensive simulations and additional data. It is conceivable that in the next few years these three sources of uncertainties will all be reduced, thus allowing Belle~II to eventually reach SM accuracy and offer an important cross check of the current exclusive anomaly.

\section{Acknowledgments}

We would like to thank Joachim Brod, Andr\'{e} Hoang, Miko\l aj Misiak and Keri Vos for helpful discussions and correspondence.
The work of JJ was supported by the Deutsche Forschungsgemeinschaft (DFG, German Research Foundation) under grant 396021762 - TRR 257. ZP acknowledges funding from the Swiss National Science Foundation (SNF) under contract 200020\_204428 and from the Slovenian Research and Innovation Agency grant No. N1-0407.

\bibliographystyle{JHEP}
\bibliography{biblio}

\end{document}